\documentclass[a4paper,11pt]{article}

\usepackage{amsfonts,amsthm,amssymb}
\usepackage{graphicx, latexsym,color, epsf}

\def\picill#1by#2(#3)
{\vbox to #2 {\hrule width #1 height 0pt depth 0pt
\vfill\epsffile{#3}}}

\newcommand{\eq}{\begin{equation}}
\newcommand{\en}{\end{equation}}
\newcommand{\eqa}{\begin{eqnarray}}
\newcommand{\ena}{\end{eqnarray}}

\begin{document}

\setlength{\unitlength}{1mm}

\thispagestyle{empty}


 \begin{center}
  { \bf  Space-Time Topology in Teleportation-Based Quantum Computation}
  \\[2mm]


{\small Yong Zhang \footnote{yong\_zhang@whu.edu.cn} and Jinglong Pang \footnote{ijpang0@gmail.com}}\\[3mm]

School of Physics and Technology, Wuhan University, P.R.China  430072\\[1cm]

\end{center}

\vspace{0.2cm}

\begin{center}
\parbox{13cm}{
\centerline{\small  \bf Abstract}  \noindent

We study the construction of both universal quantum computation and multi-partite entangled states in
the topological diagrammatical approach to quantum teleportation.  Our results
show that the teleportation-based quantum circuit model admits a space-time topological
interpretation which makes the construction of both quantum gates and four-partite entangled states
more intuitive and simpler.  Our results state
a deeper link between the space-time non-locality and the quantum non-locality which are the two-fold character
of the topological diagrammatical representation.

}

\end{center}

\vspace{.2cm}

\begin{tabbing}

\end{tabbing}

Quantum information and computation \cite{NC2011} is a newly developed interdisciplinary
field combining information science, computer science with principles of quantum mechanics.
It represents a further development of quantum mechanics, and indeed helps us to
achieve deeper understandings on quantum physics.
The quantum teleportation \cite{BBCCJPW93} is an experimentally proved quantum
information protocol instantaneously sending an unknown qubit from Alice to Bob without
any non-local physical interaction, which obliges physicists to think about again on the
relationship between Einstein's locality and the quantum non-locality \cite{NC2011, BBCCJPW93}.

The teleportation-based quantum computation \cite{GC99,Nielsen03, Leung04}
is the measurement-based quantum computation in which quantum measurement becomes the main computing
resource and determines which quantum gate is to be performed. In quantum mechanics, however, quantum
measurement breaks quantum coherence and is usually performed at the end of an experiment.  So
the measurement-based quantum computation essentially changes our conventional
viewpoint on quantum measurement and also on the standard quantum circuit model \cite{Nielsen03} in which
coherent unitary dynamics are mainly involved.
The motivation of our study on teleportation-based quantum computation is to state a deeper  link between
the space-time non-locality and the quantum non-locality, i.e., the space-time topology associated with
quantum entanglements and quantum measurements.

We apply the topological diagrammatical approach to quantum
teleportation which is developed mainly in  \cite{Kauffman05,Zhang06}.  This diagrammatical approach looks
similar to the categorical diagrammatical approach to quantum teleportation \cite{Coecke04, AC04} but they
are essentially different. In the algebraic sense, the topological approach is associated with the extension
of the Temperley--Lieb algebra \cite{TL71} which is widely used in statistical mechanics and knot theory,
see \cite{Kauffman02}, so no category theories \cite{AC04} are explicitly involved.  In the
diagrammatical sense, the categorical representation often
looks as a half of the related topological representation, which means these two approaches may give rise to
different answers on a specified question. Interested readers are invited to refer to \cite{Zhang06} for a
detailed comparison between the two diagrammatical approaches to quantum teleportation.

A single-qubit Hilbert space has an orthonormal basis vectors $|0\rangle$
and $|1\rangle$, and a single-qubit state $|\alpha\rangle$ is given by
$|\alpha\rangle= a|0\rangle +b |1\rangle$ with complex numbers $a$ and $b$.
The unit matrix $1\!\! 1_2$ and the single-qubit Pauli gates $X$ and $Z$ take the form
\eq
1\!\! 1_2 =\left(\begin{array}{cc}
              1 & 0 \\
              0 & 1 \\
            \end{array}\right), \quad X=\left(\begin{array}{cc}
              0 & 1 \\
              1 & 0 \\
            \end{array} \right), \quad  Z=\left(\begin{array}{cc}
              1 & 0 \\
              0 & -1 \\
            \end{array} \right).
\en
The four orthonormal Bell-states $|\psi(ij)\rangle, i,j=0,1$ given by
\eq
|\psi(ij)\rangle=(1\!\! 1_2 \otimes X^i Z^j) |\psi(00)\rangle, \quad
|\psi(00)\rangle =\frac 1 {\sqrt 2} (|00\rangle + |11\rangle),
\en
with the EPR state $|\psi(00)\rangle$ and with a local unitary
transformation (single-qubit gate) $X^i Z^j$, are called
 the Bell basis of the two-qubit Hilbert space.

In the topological approach \cite{Zhang06},
the Bell state $|\psi(ij)\rangle$ is represented by a cup with a solid point
denoting the single-qubit gate $X^i Z^j$,
\eqa
\setlength{\unitlength}{0.6mm}
\begin{array}{c}
\begin{picture}(45,22)
\put(2,11){$|\psi(ij)\rangle=$}
\put(32,2){\line(0,1){18}}
\put(42,2){\line(0,1){18}}
\put(32,2){\line(1,0){10}}
\put(42,11){\circle*{2.}}
\put(44.,11){\tiny{$X^i Z^j$}}
\end{picture}
\end{array}
\ena
so that a cup without a solid point denotes the EPR state $|\psi(00)\rangle$.
The complex conjugation of the Bell state, $\langle \psi(ij)|$, is represented by a cap
\eqa
\setlength{\unitlength}{0.6mm}
\begin{array}{c}
\begin{picture}(45,22)
\put(2,11){$\langle\psi(ij)|=$}
\put(32,2){\line(0,1){18}}
\put(42,2){\line(0,1){18}}
\put(32,20){\line(1,0){10}}
\put(42,11){\circle*{2.}}
\put(44.,11){\tiny{$Z^jX^i$}}
\end{picture}
\end{array}
\ena
with a solid point denoting the Hermitian conjugation of $X^i Z^j$. These diagrammatic states are called
a cup state or a cap state respectively. The projective measurements $|\psi(ij)\rangle\langle\psi(ij) |$
are called Bell measurements,
\eqa
\setlength{\unitlength}{0.6mm}
\begin{array}{c}
\begin{picture}(65,46)
\put(62,26){\line(0,1){18}}
\put(52,26){\line(0,1){18}}
\put(52,26){\line(1,0){10}}
\put(62,35){\circle*{2.}}
\put(64.,35){\tiny{$X^i Z^j$}}
\put(2,22){$|\psi(ij)\rangle \langle\psi(ij)|=$}
\put(52,2){\line(0,1){18}}
\put(62,2){\line(0,1){18}}
\put(52,20){\line(1,0){10}}
\put(62,11){\circle*{2.}}
\put(64.,11){\tiny{$Z^jX^i$}}
\end{picture}
\end{array}
\ena
represented by a top cup state with a bottom cap state.

The EPR  state $|\psi(00)\rangle$ has the nice property
\eq
(1\!\! 1_2 \otimes M)|\psi(00)\rangle =(M^T \otimes 1\!\! 1_2) |\psi(00)\rangle
\en
with $M$ denoting any single-qubit gate and the upper index $T$ denoting the
transpose conjugation. This property has a diagrammatical representation
\eqa
\setlength{\unitlength}{0.6mm}
\begin{array}{c}
\begin{picture}(45,22)
\put(2,2){\line(0,1){18}}
\put(12,2){\line(0,1){18}}
\put(2,2){\line(1,0){10}}
\put(12,11){\circle*{2.}}
\put(14.,11){\tiny{$M$}}
\put(23.,11){$=$}
\put(31.,11){\tiny{$M^T$}}
 \put(40.8,11){\circle*{2.}}
\put(40.8,2){\line(0,1){18}}
\put(50.8,2){\line(0,1){18}}
\put(40.8,2){\line(1,0){10}}
\end{picture}
\end{array}
\label{trans law: 1l}
\ena
and the similar representation also for a cap state,
\eqa
\setlength{\unitlength}{0.6mm}
\begin{array}{c}
\begin{picture}(45,22)
\put(2,2){\line(0,1){18}}
\put(12,2){\line(0,1){18}}
\put(2,20){\line(1,0){10}}
\put(12,11){\circle*{2.}}
\put(14.,11){\tiny{$M^\dag$}}
\put(23.,11){$=$}
\put(31.,11){\tiny{$M^\ast$}}
\put(40.8,11){\circle*{2.}}
\put(40.8,2){\line(0,1){18}}
\put(50.8,2){\line(0,1){18}}
\put(40.8,20){\line(1,0){10}}
\end{picture}
\end{array}
\label{trans law: 22}
\ena
with the upper index $\dag$ denoting the Hermitian conjugation and $\ast$ denoting the complex
conjugation. In the diagrammatical representations (\ref{trans law: 1l}) and (\ref{trans law: 22}),
a single-qubit gate can flow from the one branch of a cup (or cap) state to its other branch with
the transpose conjugation. This is a crucial technique in the topological operation \cite{Zhang06}  on
the following teleportation-based quantum circuits.

Let Alice and Bob share the EPR  state $|\psi(00)\rangle$, and Alice wants to transfer
an unknown quantum state $|\alpha\rangle$ to Bob, so Alice and Bob prepare the quantum
state $|\alpha\rangle\otimes |\psi(00)\rangle$, called the state preparation,
\eqa
\setlength{\unitlength}{0.6mm}
\begin{array}{c}
\begin{picture}(70,22)
\put(-2,10){$|\alpha\rangle\otimes |\psi(00)\rangle=$}
\put(45,2){\makebox(4,4){$\nabla$}}
\put(47.,6){\line(0,1){14}}
\put(57,2){\line(0,1){18}}
\put(67,2){\line(0,1){18}}
\put(57,2){\line(1,0){10}}
\end{picture}
\end{array}
\ena
in which the straight line with $\nabla$ denotes the quantum state $|\alpha\rangle$. Then, Alice performs
Bell measurements
 $|\psi(ij)\rangle\langle\psi(ij)|\otimes 1\!\! 1_2$ on the prepared state
 $|\alpha\rangle\otimes |\psi(00)\rangle$,
 \eq
 (|\psi(ij)\rangle\langle\psi(ij)|\otimes 1\!\! 1_2)(|\alpha\rangle\otimes |\psi(00)\rangle)
 =\frac 1 2 |\psi(ij)\rangle \otimes  X^i Z^j |\alpha\rangle
 \en
with the topological diagrammatical representation
\eqa
\setlength{\unitlength}{0.6mm}
\begin{array}{c}
\begin{picture}(104,80)
\put(-2,4){\vector(1,0){95}}
\put(95,4){$x$}
\put(2,0){\vector(0,1){78}}
\put(-2,76){$t$}
\put(10,12){\makebox(4,4){$\nabla$}}
\put(12.,16){\line(0,1){32}}
\put(22,39){\circle*{2.}}
\put(24,39){\tiny{$M^\dag$}}
\put(22,12){\line(0,1){36}}
\put(32,12){\line(0,1){60}}
\multiput(8,30)(1,0){30}{\line(1,0){.5}}
\put(22,12){\line(1,0){10}}
\put(12,48){\line(1,0){10}}
\put(12,54){\line(1,0){10}}
\put(12,54){\line(0,1){18}}
\put(22,54){\line(0,1){18}}
\put(22,63){\circle*{2.}}
\put(24,63){\tiny{$M$}}
\put(40, 46){\makebox(14,10){$=\frac 1 2$}}
\put(62,54){\line(1,0){10}}
\put(62,54){\line(0,1){18}}
\put(72,54){\line(0,1){18}}
\put(72,63){\circle*{2.}}
\put(74,63){\tiny{$M$}}
\put(80,12){\makebox(4,4){$\nabla$}}
\put(82.,16){\line(0,1){56}}
\put(82,39){\circle*{2.}}
\put(84,39){\tiny{$M^\ast$}}
\end{picture}
\label{tele: tl}
\end{array}
\ena
in which the single qubit gate $M=X^i Z^j$ so  $M^\ast=X^i Z^j$.  Note that the
diagram (\ref{tele: tl}) without solid points is a typical diagrammatical representation of the
Temperley--Lieb algebra  \cite{Kauffman02}.

Now let us explain the diagram (\ref{tele: tl}) in detail. The horizontal axis denotes the
space direction $x$ and the vertical axis denotes the time direction $t$, so we study the
two-dimensional space-time topology. On the left hand side of $=$, the
diagrammatical part above the dashed line denotes the Bell measurement, and the part under
the dashed line denotes the state preparation. On the right hand side of $=$, the normalization
factor $\frac 1 2$ is contributed by the normalization factors of the vanishing cup state and
cap state, and the cup state denotes the post-measurement state usually neglected in the
description of the quantum teleportation \cite{BBCCJPW93}.
The reason for $M^\ast$ is that moving $M^\dag$ from the one branch of the cup state
to the other branch leads to the transposition conjugation, $(M^{\dag})^T=M^\ast$.

Once Alice performs Bell measurements, she will tell Bob her measurement results labeled
as $(i,j)$ associated with the single-qubit gate $X^i Z^j$, then Bob will apply unitary
correction operator $Z^j X^i$ on his state to obtain the exact quantum state $|\alpha\rangle$,
\eq
 (1\!\! 1_2 \otimes 1\!\! 1_2 \otimes Z^j X^i ) (|\psi(ij)\rangle \otimes  X^i Z^j |\alpha\rangle)
  = |\psi(ij)\rangle \otimes |\alpha\rangle
\en
both of which, classical communication and unitary correction, are not shown in the
diagram (\ref{tele: tl}), for simplicity. Hence the quantum information flow sending an
unknown qubit from Alice to Bob in quantum teleportation \cite{BBCCJPW93} can be recognized as a result
of two-dimensional space-time topological operation \cite{Kauffman05, Zhang06}.

 The space-time topology in the diagrammatical teleportation (\ref{tele: tl}) may be not
that obvious. Let us consider the chained teleportation \cite{Childs03}: Alice sends an
unknown qubit $|\alpha\rangle$ to Bob with a sequence of standard teleportation protocols
\eqa
\setlength{\unitlength}{0.6mm}
\begin{array}{c}
\begin{picture}(86,51)
\put(-2,4){\vector(1,0){84}}
\put(84,4){$x$}
\put(2,0){\vector(0,1){50}}
\put(-2,48){$t$}
\put(10,12){\makebox(4,4){$\nabla$}}
\put(12.,16){\line(0,1){32}}
\put(22,12){\line(0,1){36}}
\put(32,12){\line(0,1){36}}
\multiput(8,30)(1,0){48}{\line(1,0){.5}}
\put(22,12){\line(1,0){10}}
\put(12,48){\line(1,0){10}}
\put(32,48){\line(1,0){10}}
\put(42,12){\line(0,1){36}}
\put(42,12){\line(1,0){10}}
\put(52,12){\line(0,1){36}}
\put(58,25){\makebox(14,10){$=\frac 1 4$}}
\put(72,12){\makebox(4,4){$\nabla$}}
\put(74.,16){\line(0,1){32}}
\end{picture}
\end{array}
\ena
in which the post-measurement states are neglected and only the EPR state measurements
 $|\psi(00)\rangle \langle \psi(00)|$ are considered.  The normalization factor $\frac 1 4$ is
 calculated from the normalization factors of two vanishing cup states and two vanishing cap states.
 Without unitary corrections, Bob obtain the exact quantum state $|\alpha\rangle$. Hence {\em the space-time
 topology in this paper is defined as the topological operation which straightens the configuration
 consisting of top cap states and bottom cup states}. With Bell measurements
 $|\psi(ij)\rangle\langle\psi(ij)|$, before the straightening operation, one has to move single-qubit
 gates along the path formed by top cap states with bottom cup states until boundary points of
 this path under the guidance of the properties (\ref{trans law: 1l}) and  (\ref{trans law: 22}).

An entangling two-qubit gate like the CNOT gate with single-qubit gates can perform
universal quantum computation in the quantum circuit model \cite{NC2011}. In the authors' knowledge, a
topological diagrammatical construction of two-qubit gates using teleportation \cite{GC99}
has not been done in the literature, which motivates us to study the realization of universal
quantum computation \cite{NC2011,  Barenco95b}
in the topological diagrammatical approach  \cite{Kauffman05,Zhang06}.

Quantum gates $U$  \cite{NC2011} are classified by
\eq
C_k \equiv \{ U | U C_{k-2} U^\dag \subseteq C_{k-1} \}
\en
where $C_1$ denotes the Pauli group gates, and $C_2$ denotes the Clifford group gates preserving
the Pauli group gates under conjugation.  In fault-tolerant quantum computation \cite{Gottesman97},
the $C_1$ and $C_2$ gates can be easily performed, but the $C_3$ gates may be difficultly realized.
The teleportation-based quantum computation \cite{GC99} performs $C_3$ gates by applying $C_1$ or
$C_2$ gates to preliminarily prepared quantum states with the action of $C_3$ gates.

To perform a single-qubit gate $U\in C_k$ on the unknown state $|\alpha\rangle$, Alice prepares the quantum
state $|\alpha\rangle\otimes |\psi(U)\rangle$ with $\psi(U)=(1\!\! 1_2\otimes U)|\psi(00)\rangle$, then makes
Bell measurements $|\psi(M)\rangle\langle\psi(M)|\otimes 1\!\! 1_2$ with $M=X^i Z^j$,
\eqa
\setlength{\unitlength}{0.6mm}
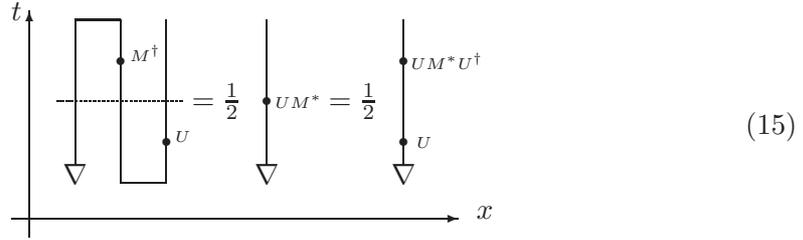
\begin{array}{c}
\begin{picture}(104,51)
\put(-2,4){\vector(1,0){98}}
\put(100,4){$x$}
\put(2,0){\vector(0,1){50}}
\put(-2,48){$t$}
\put(10,12){\makebox(4,4){$\nabla$}}
\put(12.,16){\line(0,1){32}}
\put(22,39){\circle*{2.}}
\put(24,39){\tiny{$M^\dag$}}
\put(22,12){\line(0,1){36}}
\put(32,12){\line(0,1){36}}
\put(32,21){\circle*{2.}}
\put(34,21){\tiny{$U$}}
\multiput(8,30)(1,0){28}{\line(1,0){.5}}
\put(22,12){\line(1,0){10}}
\put(12,48){\line(1,0){10}}
\put(36, 25){\makebox(14,10){$=\frac 1 2$}}
\put(52,12){\makebox(4,4){$\nabla$}}
\put(54.,16){\line(0,1){32}}
\put(54,30){\circle*{2.}}
\put(56,25){\makebox(10,10){\tiny{$U M^\ast$}}}
\put(66, 25){\makebox(14,10){$=\frac 1 2$}}
\put(82,12){\makebox(4,4){$\nabla$}}
\put(84.,16){\line(0,1){32}}
\put(84,39){\circle*{2.}}
\put(86,34){\makebox(15,10){\tiny{$U M^\ast U^\dag$}}}
\put(84,21){\circle*{2.}}
\put(86,16){\makebox(5,10){\tiny{$U$}}}
\end{picture}
\label{gate: one}
\end{array}
\ena
in which $U M^\ast=(U M^\ast U^\dag) U$. Bob performs the unitary correction operator $U M^T U^\dag \in C_{k-1}$ to
attain $U |\alpha\rangle$. Hence the difficulty of performing the single-qubit gate $U\in C_k$ becomes how to
fault-tolerantly prepare the state $|\psi(U)\rangle$ and perform the single-qubit gate $U M^T U^\dag \in C_{k-1}$.

About two-qubit gates, for examples, the CNOT gate and CZ gate \cite{NC2011},
\eqa
CNOT &=& |0\rangle\langle 0| \otimes 1\!\! 1_2 +  |1\rangle\langle 1| \otimes X, \nonumber\\
CZ &=& |0\rangle\langle 0| \otimes 1\!\! 1_2 +  |1\rangle\langle 1| \otimes Z
\ena
are both the controlled-unitary gate $CU$ and the Clifford group gate. To perform these $CU$ gates on two unknown
single-qubit states $|\alpha\rangle$ and $|\beta\rangle$, let us construct the teleportation-based quantum circuit
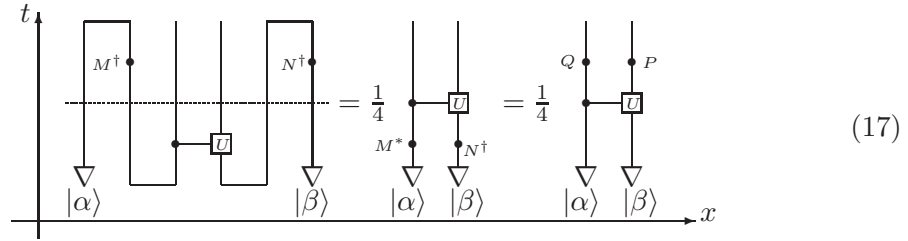
\eqa
\setlength{\unitlength}{0.6mm}
\begin{array}{c}
\begin{picture}(154,51)
\put(0,4){\vector(1,0){150}}
\put(151,4){$x$}
\put(6,0){\vector(0,1){50}}
\put(2,48){$t$}
\put(14,12){\makebox(4,4){$\nabla$}}
\put(11.5,4){\makebox(9,8){$\tiny{|\alpha\rangle}$}}
\put(16.,16){\line(0,1){32}}
\put(26,39){\circle*{2.}}
\put(18,36){\makebox(6,6){\tiny{$M^\dag$}}}
\put(26,12){\line(0,1){36}}
\put(36,12){\line(0,1){36}}
\multiput(12,30)(1,0){58}{\line(1,0){.5}}
\put(36,21){\circle*{2.}}
\put(36,21){\line(1,0){8}}
\put(44,19){\framebox(4,4){\tiny{$\,U$}}}
\put(26,12){\line(1,0){10}}
\put(16,48){\line(1,0){10}}
\put(46,12){\line(0,1){7}}
\put(46,23){\line(0,1){25}}
\put(46,12){\line(1,0){10}}
\put(56,12){\line(0,1){36}}
\put(56,48){\line(1,0){10}}
\put(66,16){\line(0,1){32}}
\put(66,39){\circle*{2.}}
\put(60,37){\makebox(4,4){\tiny{$N^\dag$}}}
\put(64,12){\makebox(4,4){$\nabla$}}
\put(61.5,4){\makebox(9,8){$\tiny{|\beta\rangle}$}}
\put(70, 25){\makebox(14,10){$=\frac 1 4$}}
\put(86,12){\makebox(4,4){$\nabla$}}
\put(82.5,4){\makebox(9,8){$\tiny{|\alpha\rangle}$}}
\put(88.,16){\line(0,1){32}}
\put(96,12){\makebox(4,4){$\nabla$}}
\put(95.5,4){\makebox(9,8){$\tiny{|\beta\rangle}$}}
\put(88,30){\circle*{2.}}
\put(88,30){\line(1,0){8}}
\put(96,28){\framebox(4,4){\tiny{$\,U$}}}
\put(98,32){\line(0,1){16}}
\put(98,16){\line(0,1){12}}
\put(88,21){\circle*{2.}}
\put(80,18){\makebox(6,6){\tiny{$M^\ast$}}}
\put(98,21){\circle*{2.}}
\put(100,18){\makebox(4,4){\tiny{$N^\dag$}}}
\put(106, 25){\makebox(14,10){$=\frac 1 4$}}
\put(124,12){\makebox(4,4){$\nabla$}}
\put(120.5,4){\makebox(9,8){$\tiny{|\alpha\rangle}$}}
\put(126.,16){\line(0,1){32}}
\put(134,12){\makebox(4,4){$\nabla$}}
\put(133.5,4){\makebox(9,8){$\tiny{|\beta\rangle}$}}
\put(126,30){\circle*{2.}}
\put(126,30){\line(1,0){8}}
\put(134,28){\framebox(4,4){\tiny{$\,U$}}}
\put(136,32){\line(0,1){16}}
\put(136,16){\line(0,1){12}}
\put(126,39){\circle*{2.}}
\put(120,37){\makebox(4,4){\tiny{$Q$}}}
\put(136,39){\circle*{2.}}
\put(138,37){\makebox(4,4){\tiny{$P$}}}
\end{picture}
\label{gate: two}
\end{array}
\ena
in which $M=X^{i_1} Z^{j_1}$ and $N=X^{i_2} Z^{j_2}$. The state preparation has the form
\eq
\label{state_prep}
(1\!\! 1_2\otimes 1\!\! 1_2\otimes CU \otimes 1\!\! 1_2\otimes 1\!\! 1_2   )
(|\alpha\rangle\otimes |\psi(00)\rangle \otimes |\psi(00)\rangle \otimes |\beta\rangle),
\en
the joint Bell measurements take the form
\eq
|\psi(M)\rangle\langle\psi(M)|\otimes 1\!\! 1_2\otimes 1\!\! 1_2 \otimes  |\psi(N)\rangle\langle\psi(N)|,
\en
then the straightening operation occurs after both moving single-qubit gates $M^\dag$ and $N^\dag$ along
the path formed by the top cat states and bottom cup states and moving the two-qubit gate $CU$ along two
vertical lines. The topological operation contributes the normalization factor $\frac 1 4$, and the
single-qubit gates $Q$ and $P$ are calculated by
\eq
CU (M^\ast \otimes N^\dag) CU^\dag =Q\otimes P
\en
where for the CNOT gate one has
\eq
Q= Z^{j_2} Z^{j_1} X^{i_1}, \quad P=X^{i_2} Z^{j_2} X^{i_1}
\en
and for the CZ gate one has
\eq
Q= Z^{i_2} X^{i_1} Z^{j_1}, \quad P=Z^{j_2} X^{i_2} Z^{i_1}
\en
and they determine which unitary correction operator, $Q^\dag\otimes P^\dag$,  to be performed in order to
attain the exact action of the Clifford gate $CU$ on the two-qubit state $|\alpha\rangle \otimes |\beta\rangle$.

Obviously, the topological construction of quantum
gates using teleportation is more intuitive and  more simpler then other original approaches
\cite{GC99,Nielsen03,Leung04}. In the topological representations, (\ref{tele: tl}), (\ref{gate: one}),
(\ref{gate: two}), (\ref{chi: one}) and (\ref{chi: two}), one can not only transport an unknown quantum state
by topological operations but also move single-qubit or two qubit gates along related configurations.

The key point
in the construction of a quantum gate $U$ using teleportation is the fault-tolerant preparation \cite{GC99}
of the multi-partite quantum state (\ref{state_prep}) with the action of this $U$ gate. In
Gottesman and Chuang's  original proposal \cite{GC99} of teleportation-based quantum computation,
a four-qubit entangled state $|\chi\rangle$ is created from two pairs of GHZ states \cite{GHZ90}. How to make
a diagrammatical representation of three-qubit entangled states such as GHZ states has been discussed in the
categorical diagrammatical approach \cite{CE11}, but we apply the quantum circuit realization \cite{NC2011} of GHZ states
so that we can clearly show how the $|\chi\rangle$ state is yielded by  a series of space-time topological operations.

The $|\chi\rangle$ state has the form
\eq
|\chi\rangle=(1\!\! 1_2 \otimes CNOT_{32} \otimes 1\!\! 1_2) (|\psi(00)\rangle \otimes |\psi(00)\rangle)
\en
with the diagrammatical representation
\eqa
\setlength{\unitlength}{0.6mm}
\begin{array}{c}
\begin{picture}(55,20)
\put(2,6){\makebox(16,10){$|\chi\rangle=$}}
\put(22,2){\line(0,1){18}}
\put(32,2){\line(0,1){18}}
\put(22,2){\line(1,0){10}}
\put(30,11){\line(1,0){12}}
\put(32,11){\circle{4.}}
\put(42,2){\line(0,1){18}}
\put(52,2){\line(0,1){18}}
\put(42,2){\line(1,0){10}}
\put(42,11){\circle*{2.}}
\end{picture}
\label{chi}
\end{array}
\ena
where the $CNOT_{ij}$ gate denotes the $i$-th qubit as the controlled qubit and the $j$-th qubit as
the target qubit.

With the Hadmard gate $H=\frac 1 2 (X+Z)$ and the CNOT gate, the EPR state $|\psi(00)\rangle$ has the
form
\eq
|\psi(00)\rangle = CNOT_{12} (H\otimes 1\!\! 1_2) |0\rangle \otimes |0\rangle
\en
with the diagrammatic representation
 \eqa
\setlength{\unitlength}{0.6mm}
\begin{array}{c}
\begin{picture}(24,22)
\put(2,6){\makebox(28,10){$|\psi(00)\rangle=$}}
\put(32,2){\makebox(4,4){$\nabla$}}
\put(34,6){\line(0,1){20}}
\put(34,10){\circle*{2.}}
\put(34,18){\circle*{2.}}
\put(37,8){\makebox(4,4){$H$}}
\put(44,2){\makebox(4,4){$\nabla$}}
\put(46,6){\line(0,1){20}}
\put(34,18){\line(1,0){14}}
\put(46,18){\circle{4}}
\end{picture}
\label{epr}
\end{array}
\ena
where the straight line with $\nabla$ denotes the state $|0\rangle$.

The three-qubit GHZ state $|G\rangle=\frac 1 2 (|000\rangle+|111\rangle)$ can be formulated by
the EPR state $|\psi(00)\rangle$ and the CNOT gate
\eq
|G\rangle=(1\!\! 1_2 \otimes CNOT_{23}) ( |\psi(00)\rangle \otimes |0\rangle )
\en
with the diagrammatical representation
 \eqa
\setlength{\unitlength}{0.6mm}
\begin{array}{c}
\begin{picture}(42,22)
\put(2,6){\makebox(16,10){$|G\rangle=$}}
\put(20,2){\line(0,1){18}}
\put(20,2){\line(1,0){10}}
\put(30,2){\line(0,1){18}}
\put(38,2){\makebox(4,4){$\nabla$}}
\put(40,6){\line(0,1){14}}
\put(30,14){\line(1,0){12}}
\put(40,14){\circle{4}}
\put(30,14){\circle*{2.}}
\end{picture}
\label{GHZ 1}
\end{array}
\ena
and so the state $(H\otimes H \otimes H) |G\rangle$ has the diagrammatical representation
 \eqa
\setlength{\unitlength}{0.6mm}
\begin{array}{c}
\begin{picture}(74,26)

\put(2,2){\line(0,1){24}}
\put(2,20){\circle*{2.}}
\put(4,18){\makebox(4,4){\tiny{$H$}}}
\put(2,2){\line(1,0){12}}
\put(14,2){\line(0,1){24}}
\put(14,20){\circle*{2.}}
\put(16,18){\makebox(4,4){\tiny{$H$}}}
\put(24,2){\makebox(4,4){$\nabla$}}
\put(26,6){\line(0,1){20}}
\put(26,20){\circle*{2.}}
\put(28,18){\makebox(4,4){\tiny{$H$}}}
\put(14,12){\line(1,0){14}}
\put(26,12){\circle{4}}
\put(14,12){\circle*{2.}}
\put(34,12){\makebox(4,4){$=$}}

\put(44,2){\line(0,1){24}}
\put(44,2){\line(1,0){12}}
\put(56,2){\line(0,1){24}}
\put(66,2){\makebox(4,4){$\nabla$}}
\put(68,6){\line(0,1){20}}
\put(54,20){\line(1,0){14}}
\put(68,20){\circle*{2.}}
\put(56,20){\circle{4}}
\put(68,12){\circle*{2.}}
\put(70,10){\makebox(4,4){\tiny{$H$}}}
\end{picture}
\label{GHZ 2}
\end{array}
\ena
where the formula $(H\otimes H) CNOT_{23} (H\otimes H) = CNOT_{32}$ is applied.

Now we are ready for the preparation of the $|\chi\rangle$ state with quantum teleportation using
GHZ states. Suppose the prior entangled six-qubit state
\eq
(H\otimes H \otimes H) |G\rangle \otimes |G\rangle,
\en
then make the Bell measurement given by
\eq
1\!\! 1_2 \otimes 1\!\! 1_2 \otimes |\psi(M)\rangle \langle \psi(M) | \otimes 1\!\! 1_2 \otimes 1\!\! 1_2
\en
with $M=X^i Z^j$, both of which give rise to the diagrammatical representation with the help of
(\ref{GHZ 1}) and (\ref{GHZ 2}),
\eqa
\setlength{\unitlength}{0.6mm}
\begin{array}{c}
\begin{picture}(154,57)
\put(0,2){\vector(1,0){154}}
\put(153,4){$x$}
\put(4,0){\vector(0,1){58}}
\put(0,54){$t$}

\multiput(6,30)(1,0){65}{\line(1,0){.5}}
\put(8,6){\line(0,1){48}}
\put(8,6){\line(1,0){12}}
\put(20,6){\line(0,1){48}}
\put(30,6){\makebox(4,4){$\nabla$}}
\put(32,10){\line(0,1){44}}
\put(18,24){\line(1,0){14}}
\put(32,24){\circle*{2.}}
\put(20,24){\circle{4}}
\put(32,16){\circle*{2.}}
\put(34,14){\makebox(4,4){\tiny{$H$}}}
\put(32,54){\line(1,0){12}}
\put(44,6){\line(0,1){48}}
\put(44,6){\line(1,0){12}}
\put(56,6){\line(0,1){48}}
\put(66,6){\makebox(4,4){$\nabla$}}
\put(68,10){\line(0,1){44}}
\put(56,42){\line(1,0){14}}
\put(68,42){\circle{4}}
\put(56,42){\circle*{2.}}
\put(44,42){\circle*{2.}}
\put(36,39){\makebox(6,6){\tiny{$M^\dag$}}}
\put(73,25){\makebox(10,8){$= \frac 1 2$}}

\put(88,6){\line(0,1){48}}
\put(88,6){\line(1,0){12}}
\put(100,6){\line(0,1){48}}
\put(98,24){\line(1,0){38}}
\put(100,24){\circle{4}}
\put(136,24){\circle*{2.}}
\put(136,10){\line(0,1){44}}
\put(134,6){\makebox(4,4){$\nabla$}}
\put(136,16){\circle*{2.}}
\put(138,14){\makebox(4,4){\tiny{$H$}}}

\put(146,6){\makebox(4,4){$\nabla$}}
\put(148,10){\line(0,1){44}}
\put(136,36){\line(1,0){14}}
\put(148,36){\circle{4}}
\put(136,36){\circle*{2.}}
\put(136,42){\circle*{2.}}
\put(139,40){\makebox(4,4){\tiny{$Z^j$}}}
\put(136,48){\circle*{2.}}
\put(139,46){\makebox(4,4){\tiny{$X^i$}}}
\put(148,48){\circle*{2.}}
\put(151,46){\makebox(4,4){\tiny{$X^i$}}}

\end{picture}
\label{chi: one}
\end{array}
\ena
where moving $M^\dag$ along the cup path and across the CNOT gate exploits  the formula
\eq
 CNOT (X^i Z^j \otimes 1\!\! 1_2) CNOT =(X^i \otimes X^i) (Z^j \otimes 1\!\! 1_2)
\en
and the vanishing top cup with the bottom cap contributes the normalization factor $\frac 1 2$.
Obviously, the $CNOT_{52}$ gate commutes with the $CNOT_{56}$ gate,  so that we continue our study
on the diagram (\ref{chi: one}) to obtain the diagram
\eqa
\setlength{\unitlength}{0.6mm}
\begin{array}{c}
\begin{picture}(154,57)
\put(0,2){\vector(1,0){154}}
\put(153,4){$x$}
\put(2,0){\vector(0,1){58}}
\put(-1,54){$t$}

\put(6,6){\line(0,1){48}}
\put(6,6){\line(1,0){12}}
\put(18,6){\line(0,1){48}}
\put(16,36){\line(1,0){38}}
\put(18,36){\circle{4}}
\put(54,24){\circle*{2.}}

\put(54,10){\line(0,1){44}}
\put(52,6){\makebox(4,4){$\nabla$}}
\put(54,16){\circle*{2.}}
\put(56,14){\makebox(4,4){\tiny{$H$}}}

\put(64,6){\makebox(4,4){$\nabla$}}
\put(66,10){\line(0,1){44}}
\put(54,24){\line(1,0){14}}
\put(66,24){\circle{4}}
\put(54,36){\circle*{2.}}

\put(54,42){\circle*{2.}}
\put(57,40){\makebox(4,4){\tiny{$Z^j$}}}
\put(54,48){\circle*{2.}}
\put(57,46){\makebox(4,4){\tiny{$X^i$}}}
\put(66,48){\circle*{2.}}
\put(69,46){\makebox(4,4){\tiny{$X^i$}}}

\put(72,28){\makebox(4,4){$=$}}

\put(82,6){\line(0,1){48}}
\put(82,6){\line(1,0){12}}
\put(94,6){\line(0,1){48}}
\put(92,36){\line(1,0){38}}
\put(94,36){\circle{4}}

\put(130,6){\line(0,1){48}}
\put(130,6){\line(1,0){12}}

\put(142,6){\line(0,1){48}}
\put(130,36){\circle*{2.}}

\put(130,42){\circle*{2.}}
\put(133,40){\makebox(4,4){\tiny{$Z^j$}}}
\put(130,48){\circle*{2.}}
\put(133,46){\makebox(4,4){\tiny{$X^i$}}}
\put(142,48){\circle*{2.}}
\put(145,46){\makebox(4,4){\tiny{$X^i$}}}

\end{picture}
\label{chi: two}
\end{array}
\ena
in which the diagrammatic representation of the EPR state (\ref{epr}) is applied.
With both classical communication and unitary correction, therefore, the four-qubit entangled
state $|\chi\rangle$ (\ref{chi}) can be exactly prepared in the diagrammatical approach.

We present the topological diagrammatical construction of both universal quantum computation
and multi-partite entangled states in this article, which is believed to represent a further development
in the diagrammatical approach to quantum information and computation \cite{Kauffman05,Zhang06,Coecke04, AC04}.
These diagrammatical representations are not usual static diagrams but dynamic by allowing topological operations.
These diagrams show that the space-time non-locality has an interpretation of the quantum non-locality associated
with quantum entanglements and quantum measurements.   What fundamental physics underlies the quantum  circuit model
is an important and open problem up to now, see Nielsen and Chuang's comprehensive comments on this problem \cite{NC2011}.
Our topological diagrammatical results clearly show that the teleportation-based quantum circuit model \cite{GC99,Nielsen03,Leung04}
can be explained as the two-dimensional space-time topological deformation of some extended Temperley--Lieb
configurations \cite{TL71, Kauffman02}. Hence the teleportation-based quantum computation
may offer us new insights on the study of the quantized space-time or quantum gravity \cite{Kauffman02}.
Teleportation-based quantum computation \cite{GC99,Nielsen03,Leung04}
is an example for measurement-based quantum computation which includes
 the one-way quantum computation \cite{RB01, BFN08, Coecke09}, so we expect that the one-way quantum computation \cite{RB01} can
 be also understood from the space-time topological viewpoint \cite{Kauffman05,Zhang06}.

 \section*{Acknowledgements}

 Yong Zhang is supported by the starting grant--273732 of Wuhan University.

\end{document}